\documentclass[11pt]{article}
\usepackage{moriond}

\def\be{\begin{equation}}
\def\ee{\end{equation}}
\def\bea{\begin{eqnarray}}
\def\eea{\end{eqnarray}}

\newcommand{\Photo}{\includegraphics[height=35mm]{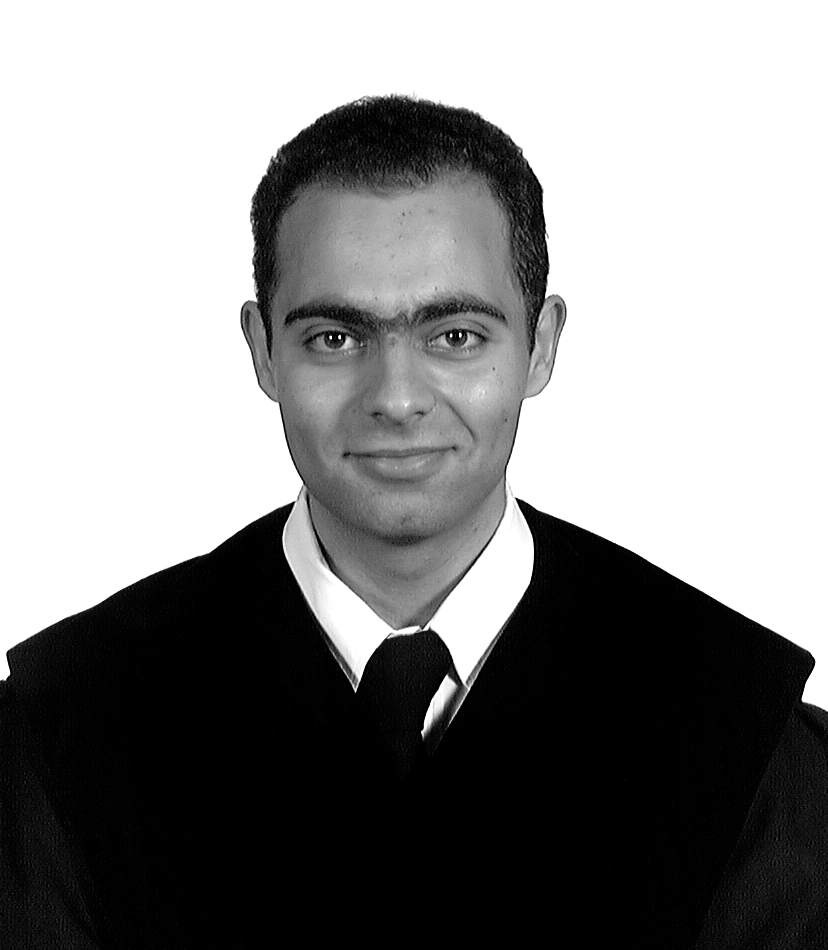}}

\begin{document}
\vspace*{4cm}
\title{$\tau^-\to\pi^-\eta^{(\prime)}\nu_\tau$ decays}

\author{PABLO ROIG}

\address{Grup de F\'{\i}sica Te\`orica, Institut de F\'{\i}sica d'Altes Energies,
Universitat Aut\`onoma de Barcelona, E-08193 Bellaterra, Barcelona, Spain}

\maketitle\abstracts{
}

\section{Introduction}
$\tau^-\to\pi^-\eta^{(\prime)}\nu_\tau$ decays belong to the so-called second-class current processes \cite{Weinberg:1958ut}: parity conservation implies that these 
decays must proceed through the vector current, which has opposed $G$-parity to the $\pi^-\eta^{(\prime)}$ system. In the limit of exact isospin symmetry $G$-parity is 
exact and these processes are forbidden. Isospin is an approximate symmetry, slightly broken both by $m_u\neq m_d$ (in QCD) and $q_u\neq q_d$ (in QED), which results 
in a sizable suppression of the considered decays, which have not been measured so far. The corresponding branching ratios upper limits are $9.9\cdot10^{-5}$ 
\cite{delAmoSanchez:2010pc} and $7.2\cdot10^{-6}$ \cite{Aubert:2008nj} and no second-class current process has been reported yet. This suppression motivates the study of 
beyond the standard model (SM) contributions to these decays \cite{Bramon:1987zb} \cite{Nussinov:2008gx}.

Here we focus on the SM prediction of these processes, focusing on the scalar and vector form factors contributions.

\section{Hadronic matrix element and decay width}
Our conventions \cite{We} are fixed from Ref.~\cite{Gasser:1984ux}. Therefore, we have ($P=\pi/\eta/\eta^\prime$)
\begin{equation}\label{Had m.e. fmas fmenos 2}
 \left\langle \pi^-P^0 \Big|\bar{d}\gamma^\mu u\Big| 0\right\rangle=c^V_{\pi^-P}\left[(p_P-p_\pi)^\mu f_+^{\pi^-P}(s)-q^\mu f_-^{\pi^-P}(s)\right]\,,
\end{equation}
with $q^\mu=(p_P+p_\pi)^\mu$, $s=q^2$, and $c^V_{\pi^-\pi^0}=-\sqrt{2}=-c^V_{\pi^-\eta^{(\prime)}}$. $f_0^{\pi^-P}(s)$, which can be used instead of $f_-^{\pi^-P}(s)$, is 
defined through
\begin{equation}\label{definition f0 2}
 \left\langle 0\Big|\partial_\mu(\bar{d}\gamma^\mu u)\Big|\pi^+P \right\rangle=i(m_d-m_u)\left\langle 0\Big|\bar{d}u\Big|\pi^+P \right\rangle\equiv i\Delta_{K^0K^+}^{QCD}c^S_{\pi^-P}f_0^{\pi^-P}(s)\,,
\end{equation}
with
\begin{equation}
c^S_{\pi^-\eta}= \sqrt{\frac{2}{3}}=c^S_{\pi^-\pi^0}\,,\quad c^S_{\pi^-\eta^\prime}= \frac{2}{\sqrt{3}}\,,\quad \Delta_{PQ}=m_P^2-m_Q^2.
\end{equation}
The mass renormalization
\begin{equation}
 m_d-m_u=\frac{\Delta_{K^0K^+}^{QCD}}{B_0}\left[1+\frac{16c_m}{F^2M_S^2}(c_d-c_m)m_K^2\right]\,,
\end{equation}
needs to be taken into account to define $f_0^{\pi^-P}(s)$.

From eqs.(\ref{Had m.e. fmas fmenos 2}) and (\ref{definition f0 2}) one gets
\begin{equation}\label{Had m.e. 2}
\left\langle \pi^-P^0 \Big|\bar{d}\gamma^\mu u\Big| 0\right\rangle=\left[(p_P-p_\pi)^\mu +
\frac{\Delta_{\pi^-P}}{s}q^\mu\right]c^V_{\pi^-P}f_+^{\pi^-P}(s)+\frac{\Delta_{K^0K^+}^{QCD}}{s}q^\mu c^S_{\pi^-P}f_0^{\pi^-P}(s)\,.
\end{equation}
The finiteness of the matrix element at the origin imposes 
\begin{equation}\label{condition origin 2}
 f_+^{\pi^-P}(0)=-\frac{c^S_{\pi^-P}}{c^V_{\pi^-P}}\frac{\Delta_{K^0K^+}^{QCD}}{\Delta_{\pi^-P}}f_0^{\pi^-P}(0)\,,
\end{equation}
which is obtained from
\begin{equation}
 f_-^{\pi^-P}(s)=-\frac{\Delta_{\pi^-P}}{s}\left[\frac{c^S_{\pi^-P}}{c^V_{\pi^-P}}\frac{\Delta_{K^0K^+}^{QCD}}{\Delta_{\pi^-P}}f_0^{\pi^-P}(s)+f_+^{\pi^-P}(s)\right]\,.
\end{equation}
In terms of these form factors, the differential decay width reads
\begin{eqnarray} \label{spectral function 2}
&& \frac{d\Gamma\left(\tau^-\to\pi^-P^0\nu_\tau\right)}{d\sqrt{s}}\,=\,\frac{G_F^2M_\tau^3}{24\pi^3s}S_{EW}\Big|V_{ud}f_+^{\pi^-P}(0)\Big|^2
\left(1-\frac{s}{M_\tau^2}\right)^2\\
& & \left\lbrace\left(1+\frac{2s}{M_\tau^2}\right)q_{\pi^-P^0}^3(s)\Big|\widetilde{f}_+^{\pi^-P}(s)\Big|^2+\frac{3\Delta_{\pi^-P^0}^2}{4s}q_{\pi^-P}(s)\Big|\widetilde{f}_0^{\pi^-P}(s)\Big|^2\right\rbrace\,,\nonumber\\
\end{eqnarray}
where
\begin{equation}
 \widetilde{f}_{+,0}^{\pi^-P}(s)=\frac{f_{+,0}^{\pi^-P}(s)}{f_{+,0}^{\pi^-P}(0)}\,,\quad q_{PQ}(s)=\frac{\lambda^{1/2}(s,m_P^2,m_Q^2)}{2\sqrt{s}}\,.
\end{equation}
Since the $\pi^-\eta^{(\prime)}$ vector form factors are proportional to the $\pi^-\pi^0$ vector form factor we may fix the first one at the origin from the latter, see 
eq.(\ref{relation pieta VFF to pipi VFF}), using that $f_+^{\pi^-\pi^0}(0)=1$. The proportionality constants will bring an overall suppression factor which explains the 
smallness of the corresponding branching fractions, in agreement with the expected vanishing in the quite accurate G-parity symmetry limit.

\section{Hadronic Form factors}
We \cite{We} have worked out the involved form factors using Chiral Perturbation Theory \cite{ChPT} including resonances within the convenient antisymmetric tensor field formalism 
\cite{RChT}, a framework which has been shown capable of providing a good description of hadronic tau decay data \cite{HadTauDec} \cite{Dumm:2013zh} \cite{Boito:2008fq}. The 
$\pi^0-\eta-\eta^\prime$ mixing has been parametrized by means of three Euler angles ($\epsilon^{\eta\pi}$, $\epsilon^{\eta^\prime\pi}$ and $\theta_{\eta\eta^\prime}$), 
including the small isospin breaking given by $z:=\frac{f_u-f_d}{f_u+f_d}$ \cite{Kroll:2005sd}. We have neglected terms of $\mathcal{O}(\epsilon^2)$ in the 
corresponding expansions.

When the vanishing of the $f_0^{\pi^-P}(s)$ form factors at large $s$ is required, one obtains the restriction $c_d=c_m=F/2$ \cite{Jamin:2000wn}, which yields
\begin{equation}\label{FFs with short distance constraints 2}
f_0^{\pi^-\pi^0}(s)=c_0^{\pi^-\pi^0}\frac{M_S^2}{M_S^2-s}\,,\quad f_0^{\pi^-\eta^{(\prime)}}(s)=c_0^{\pi^-\eta^{(\prime)}}\frac{M_S^2+\Delta_{\pi\eta^{(\prime)}}}{M_S^2-s}\,.
\end{equation}
with $c_0^{\pi^-\pi^0}=\epsilon^{\eta\pi}+\sqrt{2}\epsilon^{\eta^\prime\pi}$, $c_0^{\pi^-\eta}=\mathrm{cos}\theta_{\eta\eta^\prime}-\sqrt{2}\mathrm{sin}\theta_{\eta\eta^\prime}$, 
$c_0^{\pi^-\eta^\prime}=\mathrm{cos}\theta_{\eta\eta^\prime}+\frac{\mathrm{sin}\theta_{\eta\eta^\prime}}{\sqrt{2}}$.

We will replace $1/(M_S^2-s)$ by $1/(M_S^2-s-iM_S \Gamma_S(s))$, with the energy-dependent $a_0(980)$ width given by
\begin{equation}\label{Gamma a0}
 \Gamma_{a_0}(s)\,=\,\Gamma_{a_0}\left(M_{a_0}^2\right)\left(\frac{s}{M_{a_0}^2}\right)^{3/2}\frac{h(s)}{h\left(M_{a_0}^2\right)}\,,
\end{equation}
with
\begin{equation}\label{h(s)}
h(s)\,=\,\sigma_{KK}(s)+\frac{2}{3}\sigma_{\pi\eta}(s)\left(c_0^{\pi^-\eta}\right)^2\left(1+\frac{\Delta_{\pi\eta}}{s}\right)^2+
\frac{4}{3}\sigma_{\pi\eta^\prime}(s) \left(c_0^{\pi^-\eta^\prime}\right)^2\left(1+\frac{\Delta_{\pi\eta^\prime}}{s}\right)^2\,.
\end{equation}
In this way we are neglecting the real part of the corresponding loop functions, which will induce a small violation of analiticity (see, however, Ref.~\cite{Escribano:2010wt}).

Finally, the $\pi^-\eta^{(\prime)}$ vector form factors are obtained in terms of the well-known $\pi^-\pi^0$ vector form factor
\begin{equation}\label{relation pieta VFF to pipi VFF}
 f_{+}^{\pi^-\eta}(s)=\left[\epsilon^{\eta\pi}\mathrm{cos}\theta_{\eta\eta^\prime}-\epsilon^{\eta^\prime\pi}\mathrm{sin}\theta_{\eta\eta^\prime}\right]f_{+}^{\pi^-\pi^0}(s)\,,
\quad f_{+}^{\pi^-\eta^\prime}(s)=\left[\epsilon^{\eta^\prime\pi}\mathrm{cos}\theta_{\eta\eta^\prime}+\epsilon^{\eta\pi}\mathrm{sin}\theta_{\eta\eta^\prime}\right]f_{+}^{\pi^-\pi^0}(s)\,.
\end{equation}
Thus, we will have
\begin{equation}\label{pieta VFF at origin}
 f_{+}^{\pi^-\eta}(0)=\epsilon^{\eta\pi}\mathrm{cos}\theta_{\eta\eta^\prime}-\epsilon^{\eta^\prime\pi}\mathrm{sin}\theta_{\eta\eta^\prime}\,,\quad
f_{+}^{\pi^-\eta^\prime}(0)=\epsilon^{\eta^\prime\pi}\mathrm{cos}\theta_{\eta\eta^\prime}+\epsilon^{\eta\pi}\mathrm{sin}\theta_{\eta\eta^\prime}\,,
\end{equation}
and the normalized form factors are all the same:
\begin{equation}\label{normalized form factors}
 \widetilde{f}_{+}^{\pi^-\eta^{(\prime)}}(s)\,=\,\widetilde{f}_{+}^{\pi^-\pi^0}(s)\,,\quad \widetilde{f}_{0}^{\pi^-\pi^0}(s)\,=\,\widetilde{f}_{0}^{\pi^-\pi^0}(s)\,.
\end{equation}
While $f_{+}^{\pi^-\eta}(0)\sim\mathcal{O}(\epsilon^{\eta\pi})$, an accidental cancellation makes $f_{+}^{\pi^-\eta^\prime}(0)<\mathcal{O}\left[({\epsilon^{\eta\pi}})^2\right]$: the 
$\tau^-\to\eta\pi^-\nu_\tau$ decays are suppressed, as it corresponds to a second class current process, but the $\tau^-\to\eta^\prime\pi^-\nu_\tau$ decays are heavily 
suppressed.

\section{Phenomenological analysis}
For the vector form factor, we have taken $\widetilde{f}_{+}^{\pi^-\pi^0}(s)$ using the dispersive representation of Ref.~\cite{Dumm:2013zh} devised in Ref.~\cite{Boito:2008fq} 
for $\widetilde{f}_{+}^{K\pi}(s)$. We have estimated the model dependent error by considering Belle's data \cite{Fujikawa:2008ma} (whose extraction requires the knowledge of 
isospin-breaking corrections \cite{Cirigliano:2001er} \cite{FloresBaez:2006gf}) and the phenomenological fit made this Collaboration. This error is negligible versus the one 
coming from $\epsilon^{\eta\pi}$ and $\epsilon^{\eta^\prime\pi}$. We have fixed $\Delta_{K^0K^+}^{QCD}$ \cite{Bijnens:1993ae} \cite{Moussallam:1997xx} and determined the value 
of $z$ that fulfils eqs.(\ref{condition origin 2}) within errors \cite{We}. In this way we find $z\sim-1\cdot10^{-3}$, $\epsilon^{\eta\pi}\sim0.018(2)$ and 
$\epsilon^{\eta^\prime\pi}=5(1)\cdot10^{-3}$.

In the case of the scalar form factor the error receives important contributions both from the uncertainty on the $\epsilon^{\eta^{(\prime)}\pi}$ coefficients and on 
$M_{a_0}=(980\pm20)$ MeV and $\Gamma_{a_0}=(75\pm25)$ MeV. We have, however, neglected the contribution of a possible $a_0^\prime$ resonance, which may change sizably 
the result, especially for the $\tau^-\to\pi^-\eta^\prime\nu_\tau$ decays.

Under these assumptions we find \cite{We} $BR_+(\tau^-\to\pi^-\eta\nu_\tau)=\left(0.9\pm0.2\right)\cdot10^{-5}$, 
$BR_0(\tau^-\to\pi^-\eta\nu_\tau)=\left(2.7\pm1.1\right)\cdot10^{-5}$, which yield to $BR(\tau^-\to\pi^-\eta\nu_\tau)=(3.6\pm1.3)\cdot10^{-5}$ and 
$BR_+(\tau^-\to\pi^-\eta^\prime\nu_\tau)\in\left[10^{-11},10^{-9}\right]$, $BR_0(\tau^-\to\pi^-\eta^\prime\nu_\tau)\in\left[10^{-10},2\cdot10^{-8}\right]$, 
giving $BR(\tau^-\to\pi^-\eta^\prime\nu_\tau)\in\left[10^{-10},2\cdot10^{-8}\right]$ \footnote{Errors coming from our theoretical approach are not included. While they have 
been checked to be negligible for the vector form factor contribution, the scalar form factors in eq.~(\ref{FFs with short distance constraints 2}) need to be unitarized 
along the lines discussed in Ref.~\cite{Guo:2012yt} and, as a result, our preliminary predictions can change sizeably.}. While our predictions 
for the $\pi^-\eta$ mode are larger than previous results \cite{Nussinov:2008gx} \cite{Tisserant:1982fc} \cite{Pich:1987qq} \cite{Neufeld:1994eg} \cite{Paver:2010mz} 
\cite{Volkov:2012be} \cite{Descotes-Genon:2013uya}, our values for the $\pi^-\eta^\prime$ mode tend to be smaller \cite{Volkov:2012be} \cite{Nussinov:2009sn} 
\cite{Paver:2011md}. This is a result of our improved treatment of the $\pi^0-\eta-\eta^\prime$ mixing. We note in particular that, according to our findings, the 
$\tau^-\to\pi^-\eta\nu_\tau$ should be within discovery reach at future super-B factories.

\section*{Acknowledgments}

This work has been partly funded by the Spanish grant FPA2011-25948. Financial support of the organization covering my living expenses during the conference is acknowledged.


\section*{References}

\begin{thebibliography}{99}
\bibitem{Weinberg:1958ut}
  S.~Weinberg,
  Phys.\ Rev.\  {\bf 112} (1958) 1375.

\bibitem{delAmoSanchez:2010pc}
  P.~del Amo Sanchez {\it et al.}  [BaBar Collaboration],
  Phys.\ Rev.\ D {\bf 83} (2011) 032002.

\bibitem{Aubert:2008nj}
  B.~Aubert {\it et al.}  [BaBar Collaboration],
  Phys.\ Rev.\ D {\bf 77} (2008) 112002.

\bibitem{Bramon:1987zb}
  A.~Bramon, S.~Narison and A.~Pich,
  Phys.\ Lett.\ B {\bf 196} (1987) 543.

\bibitem{Nussinov:2008gx}
  S.~Nussinov and A.~Soffer,
  Phys.\ Rev.\ D {\bf 78} (2008) 033006.

\bibitem{We}
 R.~Escribano, S.~Gonz\'alez-Sol\'{\i}s and P.~Roig,
work in progress.

\bibitem{Gasser:1984ux}
  J.~Gasser and H.~Leutwyler,
  Nucl.\ Phys.\ B {\bf 250} (1985) 517.

\bibitem{ChPT}
  S.~Weinberg,
  Physica A {\bf 96} (1979) 327.
  J.~Gasser and H.~Leutwyler,
  Nucl.\ Phys.\ B {\bf 250} (1985) 465,
  Annals Phys.\  {\bf 158} (1984) 142.

\bibitem{RChT}
  G.~Ecker, J.~Gasser, A.~Pich and E.~de Rafael,
  Nucl.\ Phys.\ B {\bf 321} (1989) 311.
  G.~Ecker, J.~Gasser, H.~Leutwyler, A.~Pich and E.~de Rafael,
  Phys.\ Lett.\ B {\bf 223} (1989) 425.

\bibitem{HadTauDec}
  M.~Jamin, A.~Pich and J.~Portol\'es,
  Phys.\ Lett.\ B {\bf 640}, 176 (2006),
 B {\bf 640}, 176 (2006).
  D.~G.~Dumm, P.~Roig, A.~Pich and J.~Portol\'es,
  Phys.\ Rev.\ D {\bf 81} (2010) 034031,
  Phys.\ Lett.\ B {\bf 685}, 158 (2010).
  D.~R.~Boito, R.~Escribano and M.~Jamin,
  JHEP {\bf 1009}, 031 (2010).
  Z.~-H.~Guo and P.~Roig,
  Phys.\ Rev.\ D {\bf 82}, 113016 (2010).
  O.~Shekhovtsova, T.~Przedzinski, P.~Roig and Z.~Was,
  Phys.\ Rev.\ D {\bf 86}, 113008 (2012),
  O.~Shekhovtsova, I.~M.~Nugent, T.~Przedzinski, P.~Roig and Z.~Was,
  arXiv:1301.1964 [hep-ph] and 
work in progress.
  D.~G.~Dumm and P.~Roig,
  Phys.\ Rev.\ D {\bf 86}, 076009 (2012).
  A.~Guevara, G.~L\'opez~Castro and P.~Roig,
  arXiv:1306.1732 [hep-ph], to be published in Phys.\ Rev.\ D.
  R.~Escribano, S.~Gonz\'alez-Sol\'{\i}s and P.~Roig,
  arXiv:1307.7908 [hep-ph].

\bibitem{Dumm:2013zh}
  D.~G.~Dumm and P.~Roig,
  arXiv:1301.6973 [hep-ph].

\bibitem{Boito:2008fq}
  D.~R.~Boito, R.~Escribano and M.~Jamin,
  Eur.\ Phys.\ J.\ C {\bf 59} (2009) 821.

\bibitem{Kroll:2005sd}
  P.~Kroll,
  Mod.\ Phys.\ Lett.\ A {\bf 20} (2005) 2667.

\bibitem{Ambrosino:2009sc}
  F.~Ambrosino, A.~Antonelli, M.~Antonelli, F.~Archilli, P.~Beltrame, G.~Bencivenni, S.~Bertolucci and C.~Bini {\it et al.},
  JHEP {\bf 0907} (2009) 105.

\bibitem{Jamin:2000wn}
 M.~Jamin, J.~A.~Oller and A.~Pich,
  Nucl.\ Phys.\ B {\bf 587} (2000) 331,
 {\bf 622} (2002) 279,
  Phys.\ Rev.\ D {\bf 74} (2006) 074009.

\bibitem{Escribano:2010wt}
  R.~Escribano, P.~Masjuan and J.~J.~Sanz-Cillero,
  JHEP {\bf 1105} (2011) 094.

\bibitem{Fujikawa:2008ma}
  M.~Fujikawa {\it et al.}  [Belle Collaboration],
  Phys.\ Rev.\ D {\bf 78} (2008) 072006.

\bibitem{Cirigliano:2001er}
  V.~Cirigliano, G.~Ecker and H.~Neufeld,
  Phys.\ Lett.\ B {\bf 513} (2001) 361,
  JHEP {\bf 0208} (2002) 002.

\bibitem{FloresBaez:2006gf}
  F.~Flores-B\'aez, A.~Flores-Tlalpa, G.~L\'opez Castro and G.~Toledo S\'anchez,
  Phys.\ Rev.\ D {\bf 74} (2006) 071301.

\bibitem{Bijnens:1993ae}
  J.~Bijnens,
  Phys.\ Lett.\ B {\bf 306} (1993) 343.

\bibitem{Moussallam:1997xx}
  B.~Moussallam,
  Nucl.\ Phys.\ B {\bf 504} (1997) 381.

\bibitem{Guo:2012yt}
  Z.~-H.~Guo, J.~A.~Oller and J.~Ruiz de Elvira,
  Phys.\ Rev.\ D {\bf 86} (2012) 054006.

\bibitem{Tisserant:1982fc}
  S.~Tisserant, and T.~N.~Truong,
  Phys.\ Lett.\ B {\bf 115} (1982) 264.

 \bibitem{Pich:1987qq}
  A.~Pich,
  Phys.\ Lett.\ B {\bf 196} (1987) 561.

\bibitem{Neufeld:1994eg}
  H.~Neufeld and H.~Rupertsberger,
  Z.\ Phys.\ C {\bf 68} (1995) 91.

\bibitem{Paver:2010mz}
  N.~Paver and Riazuddin,
  Phys.\ Rev.\ D {\bf 82} (2010) 057301.

\bibitem{Volkov:2012be}
  M.~K.~Volkov and D.~G.~Kostunin,
  Phys.\ Rev.\ D {\bf 86} (2012) 013005.

\bibitem{Descotes-Genon:2013uya}
  S.~Descotes-Genon, E.~Kou, and B.~Moussallam,
  arXiv:1303.2879 [hep-ph].

\bibitem{Nussinov:2009sn}
  S.~Nussinov and A.~Soffer,
  Phys.\ Rev.\ D {\bf 80} (2009) 033010.

\bibitem{Paver:2011md}
  N.~Paver and Riazuddin,
  Phys.\ Rev.\ D {\bf 84} (2011) 017302.
\end{thebibliography}
\end{document}